\documentstyle[11pt,newpasp,twoside]{article}
\def\edcomment#1{\iffalse\marginpar{\raggedright\sl#1\/}\else\relax\fi}
\reversemarginpar

\begin{document}
\title{X-ray Emission \\ from Old and Intermediate Age Brown Dwarfs}
 \author{Beate Stelzer \& Ralph Neuh\"auser}
\affil{Max-Planck-Institut f\"ur extraterrestrische Physik, Postfach 1312, D-85741 Garching, Germany}

\begin{abstract}
We report on two recent {\em XMM-Newton} observations of Brown Dwarfs in the Pleiades cluster and in the field aiming to constrain the age dependence of X-ray emission from substellar objects.
\end{abstract}

\section{Introduction}

The standard picture of solar-type magnetic activity 
is expected to break down for very-low mass stars: being fully
convective throughout the interior they lack the interface between radiative
core and convective envelope in which the solar-type $\alpha\Omega$-dynamo 
is thought to reside. 
In spite of these theoretical predictions X-ray and H$\alpha$ activity 
has been observed on stars with masses below the fully convective boundary,
corresponding to spectral type $\sim$\,M3 
(Fleming et~al. 1995, Gizis et~al. 2000). Recent H$\alpha$ observations 
of ultracool field dwarfs indicate, however, a decline of activity
setting in near the substellar limit at 
spectral type M9 (Mohanty \& Basri 2002, see also Basri this volume).

While there is a substantial data base on chromospheric activity, only few 
X-ray observations of very low-mass (VLM) field dwarfs 
have been performed so far. Virtually all of the
X-ray emitting field dwarfs with spectral type later than $\sim$ M7 
have been detected 
only during a temporary outburst, with quiescent emission below the detection 
threshold (see Sect.~3). Whether this is due to a lack of sensitivity of the 
respective observations, or whether these objects indeed are X-ray quiet
can now be tested with a new generation of X-ray instruments onboard
{\em XMM-Newton} and {\em Chandra}.

X-ray observations with {\em ROSAT} have shown that young 
Brown Dwarfs (BDs) are more readily 
detected than older ones, as they show higher 
levels of activity (Neuh\"auser et~al. 1999). Due to the absence of an
internal energy source 
the evolution of BDs goes along with a decrease of 
effective temperature. The accompanying drop of the ionization fraction
may prevent coupling of the gas to the magnetic field, 
thus shutting off activity. 
Probing the relation between activity and the evolution of 
atmospheric conditions requires high-sensitivity X-ray observations 
of VLM stars and BDs at different ages.

\section{Observations}\label{sect:observations}

In order to examine the dependence of X-ray emission on 
age and/or effective temperature, we observed two 
BDs in different evolutionary stages: 
(a) the $\sim 100$\,Myr old BD CFHT-Pl\,12 in the Pleiades cluster, 
and (b) Denis\,J1228-1547 which is a BD binary in the field at an age 
of $\sim 500$\,Myr. 

The Pleiades cluster has been extensively monitored with past and present
day X-ray instrumentation (see sky map in Fig.~1 presenting
all X-ray observations centered on the Pleiades since the {\em ROSAT}
mission). 
None of the earlier observations was deep enough to reach into the 
substellar regime. In particular
the limiting sensitivity of pointed {\em ROSAT} observations in the
Pleiades region is $\lg{L_{\rm x}} \sim 28.5 ... 29.0$~erg/s 
(Stelzer \& Neuh\"auser 2001). 
Two long exposures
with {\em Chandra} included no BDs in the field (Krishnamurthi et~al. 2000, 
Daniel et~al. 2002). Here, we discuss
one of the three {\em XMM-Newton} pointings performed up to now. 
\begin{figure}
\begin{center}
\plotfiddle{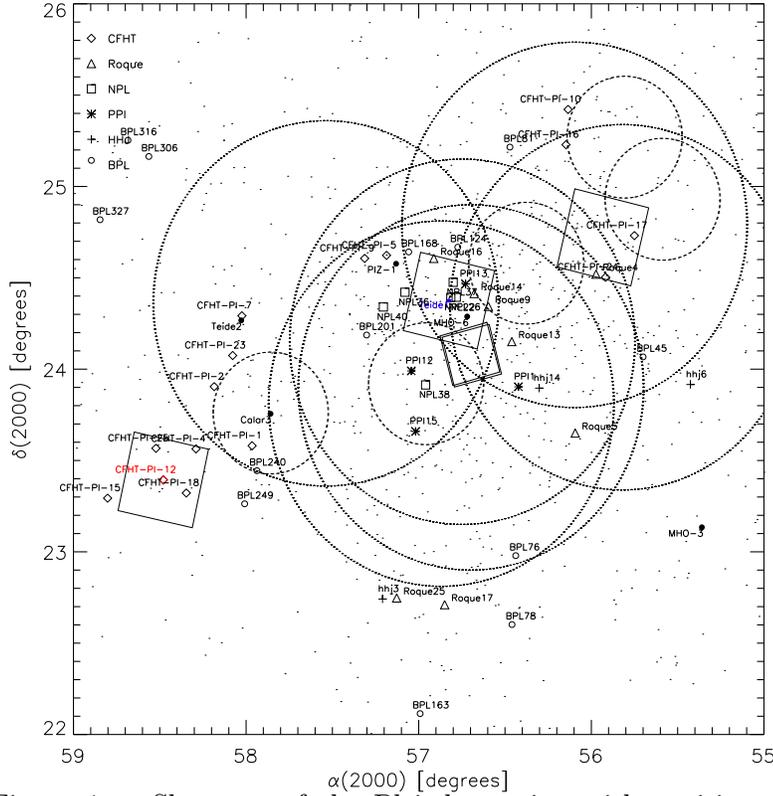}{10cm}{0}{70}{70}{-180}{-20}
\caption{Sky map of the Pleiades region with position of X-ray pointings by {\em ROSAT} (small circles = HRI, large circles = PSPC), {\em Chandra} (small squares), and {\em XMM-Newton} (large squares); dots are Pleiades members.} 
\end{center}
\end{figure}
As seen from Fig.~1 CFHT-Pl\,12 is located in a region of the cluster
which is widely unexplored in X-rays so far. CFHT-Pl\,12 is the optically
brightest of the BD candidates in the Pleiades, and its position above the 
main-sequence in the color-magnitude diagram indicates it could be a binary
(Bouvier et~al. 1998).
Unresolved binaries should have a higher probability of detection, 
and therefore also 
Denis\,J1228-1547, a BD binary with $\sim 0.3^{\prime\prime}$ separation, 
is a favorable target for X-ray studies.

\section{Results}

The data were analysed using the standard {\em XMM-Newton} SAS, version 5.3.0.
Strong background flaring restricted the useful exposure time to 
$\sim 8.5$\,ksec for CFHT-Pl\,12 and $\sim 6$\,ksec for Denis\,J1228-1547.
Source detection was performed in three energy bands in the range 
from $0.3$ to $5.0$\,keV using a combination of local, map, and
maximum likelihood detection mechanism. Neither CFHT-Pl\,12 nor
Denis\,J1228-1547 are detected at a $ML$ detection threshold of $10$.
Upper limits to the source flux derived under the assumption of
a one-temperature, 1\,keV Raymond-Smith spectrum plus photo-absorption
are listed in Table~1. To facilitate comparison with earlier
X-ray observations of VLM objects, the luminosities were converted
to the $0.1-2.4$\,keV band covered by {\em ROSAT}.
\begin{table}
\begin{center}
\caption{X-ray properties of two BDs observed with {\em XMM-Newton}. Counts are measured in the broad band of EPIC ($0.3-5$\,keV), but X-ray flux and luminosity and their 3$\sigma$ confidence upper limits apply to the {\em ROSAT} band of $0.1-2.4$\,keV; last column are data from RASS.}
\label{tab:results}
\begin{tabular}{lrrrrrr} \hline
Instr.  & Cts    & Expo   & \multicolumn{1}{c}{$F_{\rm x}$} & $\lg{L_{\rm x}}$ & $\lg{(\frac{L_{\rm x}}{L_{\rm bol}})}$ & $\lg{L_{\rm x}}$ \\
             &        & \multicolumn{1}{c}{[s]} & [${\rm erg/s/cm^2}$] & [erg/s] & & (RASS) \\ 
\hline \hline 
\multicolumn{6}{c}{CFHT-Pl\,12} & $< 29.3$ \\ \hline
pn      & $14$   & $~8593$ & $< 4.1 \cdot 10^{-15}$ & $< 27.28$ & $< -3.48$ & \\
MOS\,1+2& $29$   & $34481$ & $< 3.5 \cdot 10^{-15}$ & $< 27.76$ & $< -3.00$ & \\
\hline \hline
\multicolumn{6}{c}{Denis\,J1228-1547} & $< 27.3$ \\ \hline
pn      & $20$   & $5889$ & $< 4.0 \cdot 10^{-15}$  & $< 26.15$ & $< -3.14$ & \\
MOS\,1+2 & $15$ & $15596$ & $< 5.8 \cdot 10^{-15}$ & $< 26.30$ & $< -2.99$  & \\
\hline
\end{tabular}
\end{center}
\end{table}

In Fig.~2 the newly derived upper limit for Denis\,J1228-1547 is compared to
X-ray observations of other field dwarfs: 
M-stars from the Catalog of Nearby
Stars (CNS; Gliese \& Jahreiss 1991) detected in the {\em ROSAT} All-Sky Survey
(=RASS; data from H\"unsch et~al. 1999), and 
field dwarfs later than M6 observed with {\em ROSAT} in pointed observations
and/or with {\em Chandra} (see Neuh\"auser et~al. 1999, 
Fleming et~al. 2000, Rutledge et~al. 2000, Schmitt \& Liefke 2002).
The {\em XMM-Newton} observation of Denis\,J1228-1547 presented here is the 
first attempt to extend the study of X-ray activity into spectral class L.
The upper limit derived from the RASS is improved by $\sim$ 1 order of 
magnitude for this object
despite the considerable loss of observing time (due to high background).
Note, that at the time of target selection only three 
BDs in the field had been discovered. In the near future, 
longer {\em XMM-Newton} observations of the nearby early L-type
dwarfs identified since then 
are likely to provide faint detections or meaningful upper limits
populating the lower right of Fig.~2.

\begin{figure}[t]
\begin{center}
\plotfiddle{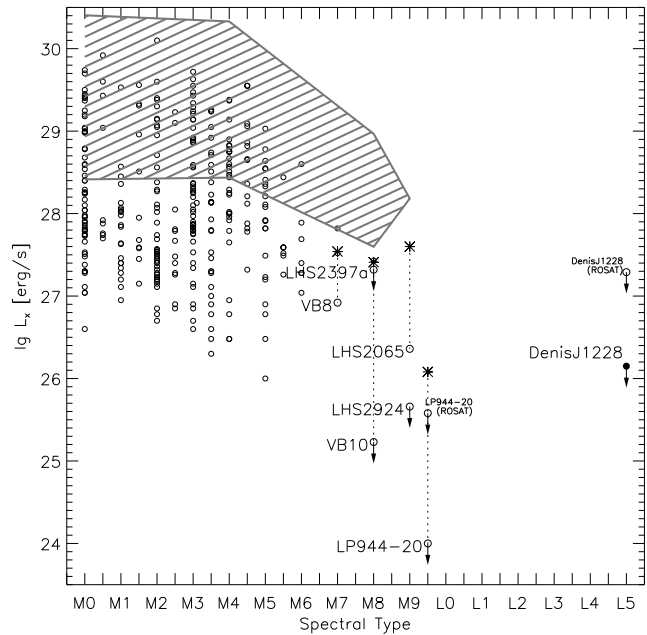}{7.6cm}{0}{60}{60}{-145}{-15}
\caption{$\lg{L_{\rm x}}$ over spectral type for field M and L dwarfs. Asterisks denote flares, arrows denote upper limits for non-detections. For Denis\,J1228-1547 and LP\,944-20 both the {\em ROSAT} and the {\em XMM-Newton} or {\em Chandra} measurements are given. The typical region occupied by younger objects of the same spectral type found in star forming regions (data from Mokler \& Stelzer 2002) is shown as hatched area.}
\end{center}
\end{figure}

\end{document}